# Landau diamagnetism within nonextensive statistical thermodynamics

İsmail Sökmen[a], Fevzi Büyükkılıç[b†], Doğan Demirhan[b]

[a] *Department of Physics, Faculty of Arts&Sciences, Dokuz Eylül University, Alsancak İzmir, Turkey*

[b] *Department of Physics, Faculty of Sciences, Ege University, 35100 Bornova İzmir, Turkey*

**Abstract:**

In this paper, Landau diamagnetism is revisited in the context of nonextensive statistical thermodynamics. Exact as well as approximate expressions for partition function, magnetization and susceptibility are obtained by using Hilhorst integral transformation which provides exact solutions in the framework of Tsallis statistics. The results have been reassessed with the standard results to illustrate the effect of nonextensivity.

___

[†]e-mail:fevzi@fenfak.ege.edu.tr

# 1. Introduction

It has been understood that extensive (additive) Boltzmann-Gibbs (BG) thermal statistics fails to study the nonextensive physical systems where long-range interactions or long-range microscopic memory are involved or, the system evolves in a (multi)fractal space -time. Thus, standard statistical mechanics, consequently, standard thermodynamics is not universal and is suitable for extensive systems. A generalized formalism is proposed by C. Tsallis which might overcome some of the difficulties that nonextensive physical systems exhibit [1,2,3].

This generalization relies on a new entropic form for the entropy that is inspired from (multi)fractals

$$S_q = k_B \frac{1 - \sum_{i=1}^{W} p_i^q}{q-1} \qquad (q \in R)$$

where $k_B$ is a positive constant and W is the total number of microscopic accessible states of the system (for the q<0 case, those probabilities which are not positive should be excluded). This expression recovers the well-known standard Shannon entropy in the limit q→1.

$$S_1 = -k_B \sum_{i=1}^{W} p_i \ln p_i$$

The entropic index q, related to and determined by the microscopic dynamics, characterizes the degree of nonextensivity.

Recently, generalized statistical mechanics has been successfully applied to investigate physical systems which exhibits nonextensive features. Amongst them, there are stellar polytrops [3], Levy-like anomalous diffusions [4], two dimensional turbulence[5], solar neutrino problem [6] velocity distributions of galaxy clusters[7],

Cosmic background radiation[8,9] and correlated themes[10], linear response theory[11], thermalization of electron-phonon system[12] and low dimensional dissipative systems[13].

It should be remarked that what makes $S_q$ favourable is that it has, with regard to $\{p_i\}$, definite concavity property for all values of q[14].

In order to have a sensible results which are exact or approximate some techniques of calculations have been formulated in Tsallis generalized statistical mechanics formalism (TT). These are (1-q) expansion [8], factorization method for quantal distribution functions [15], perturbative expansion [16], variational methods [16,17], semiclassical approximation [18], Feynman's path integrals method in nonextensive physics [19], and Green functions [20].

In this paper however, we have revisited the Landau diamagnetism where it is investigated by incorporating coherent states and using factorization method within TT[21]. Here, our motivation is to redrive the physical quantities such as generalized partition function, magnetization and susceptibility exactly as well as approximately by incorporating Hilhorst Integral transformation in the context of nonextensive thermostatistics, moreover, to reassess the obtained results with the standard ones. We believe that our approach will help to understand the difference between a perfect conductor and a superconductor in terms of entropic index q.

## 2. Energy states of an electron in a uniform magnetic field and diamagnetism

Diamagnetism arises from the quantization of the induced circular current of charged particles in an external magnetic field. According to the Lénz rule, the induced dipole moments oppose to the direction of the applied field, thus the diamagnetic materials weaken it and displacement of the magnetic field out of a diamagnetic material is observed. In case of ideal diamagnetic material such as superconductors the interior of the materials is free of the applied field, this effect is known as Meissner-Ochsenfeld effect more commonly referred to as Meissner effect (Berlin,1933).

Here, the magnetic behaviour of the electrons has been considered. Permanent magnetic dipole moments of electrons and nuclei are ignored and only the influence of the external magnetic fields on the electrons is taken into account.

The Hamiltonian of a nonrelativistic free electron in an external magnetic field is

$$\mathcal{H} = \frac{1}{2m}\left(\vec{P} + \frac{e}{c}\vec{A}\right)^2 \tag{1}$$

where e is positive. The Schrödinger equation

$$\mathcal{H}\Psi = \varepsilon\psi \tag{2}$$

is invariant under the Coulomb gauge transformation where $\vec{\Delta}\cdot\vec{A} = 0$. We consider a uniform external magnetic field H along the z-axis, and choose the vector potential

$$\vec{A} = \frac{1}{2m}\left(-\frac{1}{2}H_y, \frac{1}{2}H_x, 0\right) \tag{3}$$

so that $\vec{H} = H\vec{k}$, thus the Hamiltonian of an electron is written down as

$$\mathcal{H} = \frac{1}{2m}\left[\left(P_x - \frac{1}{z}m\omega y\right)^2 + \left(P_y + \frac{1}{2}m\omega x\right)^2 + P_z^2\right] \tag{4}$$

where $\omega = \frac{eH}{mc}$ is the electron cyclotron frequency m is the mass of electron, and c is the velocity of light. Introducing the operators

$$\pi_\pm = P_x \pm iP_y \pm \left(\frac{i\hbar}{z\ell^2}\right)(x \pm iy) \tag{5}$$

where $\ell = \left(\dfrac{\hbar}{m\omega}\right)^{1/2}$ is the classical radius of ground state Landau orbit and noting that they obey the commutation relation

$$[\pi_-, \pi_+] = 2m\hbar\omega \tag{6}$$

It can be shown that the Hamiltonian becomes:

$$\mathcal{H} = \dfrac{P_z^2}{2m} + \left(\dfrac{\pi_+ \pi_-}{2m}\right) + \dfrac{1}{2}\hbar\omega. \tag{7}$$

Eq.(7) can be separated into parallel and transverse components: Thus, the solution of the Schrödinger Equation of the Hamiltonian which is given by Eq.(2) consists of the motion along the z direction of a free particle whose Hamiltonian $\mathcal{H}_z$ and the motion of harmonic oscillator whose Hamiltonian $\mathcal{H}_t$ and move on the xy plane

$$\mathcal{H} = \mathcal{H}_z + \mathcal{H}_t \tag{8}$$

where

$$\mathcal{H}_z = \dfrac{P_z^2}{2m}. \tag{9}$$

The energy of free particle is known. On the other hand, choosing the convenient basis of set, for the harmonic oscillator the energy eigenvalues are obtained:

$$\varepsilon_n = \dfrac{eH}{mc}\left(n + \dfrac{1}{2}\right) \qquad \text{n=0,1,2,} \tag{10}$$

The energy levels of Eq. (10) are degenerate. The number of states g which belong to a discrete n are calculated [26]. The levels of the field-free case which superpose have an energy $\varepsilon = \dfrac{1}{2m}\left(P_x^2 + P_y^2\right)$ which lies between the energy interval $2\mu_B H n$ and $2\mu_B H(n+1)$. Thus

$$g = \dfrac{1}{h^2}\int dP_x dP_y dx dy \tag{11}$$

$$2\mu_B H n \leq \varepsilon \leq 2\mu_B H(n+1)$$

with

$$\varepsilon = \frac{1}{2m}\left(P_x^2 + P_y^2\right), \; \mu_B = \frac{e\hbar}{2mc} \tag{12}$$

The integral over x and y just yields the base area of the container, the momentum integrals can be solved by substituting plane polar coordinates:

$$g = \frac{S}{h^2} 2\pi \int_{P_n}^{P_{n+1}} P dP = \frac{S}{h^2}\pi\left(P_{n+1}^2 - P_n^2\right)$$

or

$$g = S\frac{eH}{hc} \tag{13}$$

where $P_n = \sqrt{4m\mu_B H n}$.

Degeneracy is independent of n. It vanishes while H→0, since we have assumed the limiting case of a continuos spectrum in Eq. (12). In our case we take a cylindrical sample of length $L_z$, radius R, therefore base area is $S=\pi R^2$ and it is oriented along the magnetic field H.

We consider a system, which is in a thermodynamical equilibrium state, and fluctuate in terms of energy with its environment, the generalized partition function for the canonical ensemble within the TT formalism is given by [1,2].

$$Z_q = Tr\left\{[1-(1-q)\beta\mathcal{H}]^{\frac{1}{1-q}}\right\} \tag{14}$$

where $\beta = \frac{1}{k_B T}$. The generalized free energy $F_q$ of the sample is of which the electron density n ,is obtained from the generalized partition function $Z_q$,

$$F_q = -\frac{n}{\beta}\frac{Z_q^{1-q}-1}{1-q} \tag{15}$$

and generalized magnetization $M_q$ is

$$M_q = -\frac{\partial F_q}{\partial H} \tag{16}$$

and generalized susceptibility is

$$\chi_q = -\frac{\partial M_q}{\partial H}. \tag{17}$$

### 3. Hilhorst Integral Transformation and Generalized Canonical Partition Function For Landau Diamagnetism

Hilhorst integral transformation [22] and extension for q<1 bridges the generalized statistical mechanics to its respective standard quantities thus generalized thermodynamics could easily be established [23]. Therefore the generalized statistical mechanics can be regarded as a Hilhorst integral transformation of respective standart quantities. From the definition of Gamma function [24]

$$\Gamma(\alpha) = \int_0^\infty dt\, e^{-t} t^{\alpha-1} \tag{18}$$

substituting

$$t = \upsilon[1-(1-q)\beta\mathcal{H}] \quad\text{and}\quad \alpha = \frac{1}{1-q} \quad\text{into Eq.(18)}$$

one obtains:

$$[1-(1-q)\beta\mathcal{H}]^{\frac{1}{1-q}} = \frac{1}{\Gamma\left(\frac{1}{q-1}\right)}\int_0^\infty d\upsilon\, \upsilon^{\frac{1}{q-1}-1}\exp[-\upsilon(q-1)\beta\mathcal{H}] \tag{19}$$

On the other hand, the Trace of Eq.(19) provides the generalized Canonical-partition function $Z_q$ for Landau Diamagnetism, which can be evaluated by choosing a convenient set of basis states:

$$Z_q(\beta) = \frac{1}{\Gamma\left(\frac{1}{q-1}\right)} \int_0^\infty d\upsilon \, \upsilon^{\frac{1}{q-1}-1} e^{-\upsilon} Z_1(\upsilon(q-1)\beta) \tag{20}$$

where

$$Z_1(\upsilon(q-1)\beta) = Tr\left\{e^{-\upsilon(q-1)\beta H}\right\}. \tag{21}$$

Substituting the components of the Hamiltonian of Eq.(9) into Eq.(14), the partition function $Z_1(\upsilon(q-1)\beta)$ is separated in two parts

$$Z_1(\upsilon(q-1)\beta) = Z_{1//}(\upsilon(q-1)\beta) Z_{1\perp}(\upsilon(q-1)\beta) \tag{22}$$

where $Z_{1//}$ is the partition function, which is parallel to the cylinder,

$$Z_{1//}(\upsilon(q-1)\beta) = L_z \left[\frac{m}{2\pi\hbar^2 \beta \upsilon(q-1)}\right]^{\frac{1}{2}} \tag{23}$$

where $L_z$ is the height of the cylinder. On the other hand the transverse part is

$$Z_{1\perp}(\upsilon(q-1)\beta) = g \frac{\exp(-\upsilon(q-1)x)}{1-\exp(-\upsilon(q-1)x)} \tag{24}$$

where g is the number of the states which is given by Eq.(13) and $x = \beta\hbar \frac{eH}{mc}$. Using Eq.(13) for the geometry of the cylinder the number of the states is

$$g = 2\pi R^2 \frac{eH}{hc}.$$

By substituting of Eq.(23) and Eq.(24) into Eq.(22) one obtains

$$Z_q(\beta) = V\left(\frac{m\omega}{2\pi\hbar}\right)\left(\frac{m}{2\pi\hbar^2}\right)^{1/2} \frac{1}{(q-1)^{1/2}\Gamma\left(\frac{1}{q-1}\right)} I_q \qquad (25)$$

where

$$I_q = \int_0^\infty d\upsilon\, \upsilon^{\left(\frac{1}{q-1}-\frac{1}{2}\right)-1} e^{-\upsilon} \frac{\exp\left(-\upsilon(q-1)x/2\right)}{[1-\exp(-\upsilon(q-1)x)]}. \qquad (26)$$

One can write Eq.[26] as

$$I_q = \sum_{n=0}^\infty \int_0^\infty d\upsilon\, \upsilon^{\left(\frac{1}{q-1}-\frac{1}{2}\right)-1} \exp\left(-\upsilon\left[1-(1-q)x(n+1/2)\right]\right) . \qquad (27)$$

In order to terminate the integral transformation, one may, once more, recourse to the change of integral variable $\upsilon$ in Eq.(27)

$$u = \upsilon\left[1-(1-q)x\left(n+\frac{1}{2}\right)\right] \qquad (28)$$

then Eq.(27) reads,

$$I_q = \Gamma\left(\frac{1}{q-1}-\frac{1}{2}\right) \sum_{n=0}^\infty \left[1-(1-q)x\left(n+\frac{1}{2}\right)\right]^{\frac{1}{1-q}+\frac{1}{2}}. \qquad (29)$$

Since $\frac{1}{q-1}-\frac{1}{2} \geq 0$ one can draw the conclusion $1 \leq q \leq 3$ which is in agreement with the distribution function of paramagnetism where a fractal approach is followed [25]. Finally, Eq.(25) becomes

$$Z_q(\beta) = V\left(\frac{m\omega}{2\pi\hbar}\right)\left(\frac{m}{2\pi\hbar^2}\right)^{1/2} \Gamma(a,b) \sum_{n=0}^\infty \left[1-(1-q)x\left(n+\frac{1}{2}\right)\right]^{\frac{1}{1-q}+\frac{1}{2}} \qquad (30)$$

where $\quad \Gamma(a,b) = \frac{\Gamma(a-b)}{\Gamma(a)} e^{b\log a} \qquad a = \frac{1}{q-1}, b = \frac{1}{2}.$

When a→∞ i.e. q →1 it is easy to show that $Z_q(\beta)$ turns to the standard partition function, i.e.

$$Z_1(\beta) = V\left(\frac{m\omega}{2\pi\hbar}\right)\left(\frac{m}{2\pi\hbar^2\beta}\right)^{1/2} \frac{e^{-x/2}}{1-e^{-x}} \qquad (31)$$

since

$$\lim_{a\to\infty} \Gamma(a,b) \to 1 \quad [27].$$

In the case of high temperature i.e $x = \beta\hbar\omega \ll 1$ generalized partition function simplifies and Eq.(30) becomes

$$Z_q(\beta) = V\frac{m\omega}{2\pi\hbar}\left(\frac{m}{2\pi\hbar^2}\right)^{1/2} \Gamma(a,b) \frac{[1-(1-q)x]^{-\frac{1}{q-1}+\frac{1}{2}}}{1-[1-(1-q)x]^{-\frac{1}{q-1}+\frac{1}{2}}} \qquad q>1 \qquad (32)$$

since $1-(1-q)x\left(n+\frac{1}{2}\right) \approx \exp\left[-(1-q)x\left(n+\frac{1}{2}\right)\right]$

including zero-point energy.

Now, Let us obtain the generalized partition function for the case q<1. The contour integral, which is suitable for q<1 case is [ 27 ]

$$\Gamma(\alpha)\frac{i}{2\pi}\oint_c dt(-t)^{-\alpha}e^{-t} = 1 \qquad (33a)$$

where α>0 .

Substitution of

$$t = \upsilon[1-(1-q)\beta\mathcal{H}]; \quad \text{and} \quad \alpha = 1+\frac{1}{1-q} \quad (\alpha > 0 \text{ since q<1}) \qquad (33b)$$

into Eq.(33a) leads to:

$$[1-(1-q)\beta\mathcal{H}]^{\frac{1}{1-q}} = \frac{i}{2\pi}\Gamma\left(\frac{2-q}{1-q}\right)\oint_C dv(-v)^{-1-\frac{1}{1-q}}\exp(-v)\exp(v(1-q)\beta\mathcal{H}). \quad (34)$$

In order to obtain the generalized partition function from Eq.(34) a convenient set of basis states are chosen, then the trace of the corresponding matrix is taken. Thus, the partition function of the nonextensive system is obtained for q<1:

$$Z_q(\beta) = \Gamma\left(\frac{2-q}{1-q}\right)\frac{i}{2\pi}\oint_C dv(-v)^{\frac{1}{1-q}-1}\exp(-v)Z_1(-\beta v(1-q)). \quad (35)$$

Similar to the arguments which is given above for the q>1 case, the components of $Z_1(-\beta v(1-q))$ consists of two parts one is parallel other is transverse to the axis of the sample. Thus,

$$Z_1(-\beta v(1-q)) = Z_{1//}(-v(1-q)\beta)Z_{1\perp}(-v(1-q)\beta) \quad (36)$$

where

$$Z_{1//}(-v(1-q)\beta) = L_z\left(\frac{m}{2\pi\hbar^2(-v(1-q)\beta)}\right)^{1/2} = L_z\left(\frac{m}{2\pi\hbar^2\beta}\right)^{1/2}\frac{1}{(-v)^{1/2}(1-q)^{1/2}} \quad (37)$$

and

$$Z_{1\perp}(-v(1-q\beta)) = g\frac{\exp(-v(1-q)x/2)}{1-\exp(-v(1-q)x)}. \quad (38)$$

The number of the degenerate states g in Eq.(38) is given by Eq.(13).

By substituting Eq.(37) and Eq.(38) into Eq.(36) the partition function which given by Eq.(36) is found. When $Z_1(-\beta v(1-q))$ given by Eq.(36) is replaced in Eq.(35) $Z_q(\beta)$ is obtaied as :

$$Z_q(\beta) = V\left(\frac{m\omega}{2\pi\hbar}\right)\left(\frac{m}{2\pi\hbar\beta}\right)^{1/2} \frac{\Gamma\left(\frac{1}{1-q}+1\right)}{(1-q)^{1/2}\Gamma\left(\frac{1}{1-q}+\frac{3}{2}\right)} \sum_{n=0}^{\infty}\left[1-(1-q)x\left(n+\frac{1}{2}\right)\right]^{\frac{1}{1-q}+\frac{1}{2}}. \quad (39)$$

Finally, in view of Eq.(30) and Eq.(39) $Z_q(\beta)$ is written for different q's:

$$Z_q(\beta) = V\left(\frac{m}{2\pi\hbar^2\beta}\right)^{3/2} xX_q \begin{cases} \dfrac{\Gamma\left(\frac{1}{q-1}-\frac{1}{2}\right)}{(q-1)^{1/2}\Gamma\left(\frac{1}{q-1}\right)} & q > 1 \\[2ex] \dfrac{\Gamma\left(\frac{1}{1-q}+1\right)}{(1-q)^{1/2}\Gamma\left(\frac{1}{1-q}+\frac{3}{2}\right)} & q < 1 \end{cases} \quad (40)$$

where

$$X_q = \sum_{n=0}^{\infty}\left[1-(1-q)\left(n+\frac{1}{2}\right)x\right]^{\frac{1}{1-q}+\frac{1}{2}}. \quad (41)$$

It is wise to take the logarithm of $Z_q(\beta)$ and its derivative with respect to x since one needs them for advancing further:

$$\ln Z_q = \ln C + \ln x + \ln X_q \quad (42)$$

where C stands for all constants and

$$\frac{\partial}{\partial x}\ln Z_q = \frac{1}{x} + \frac{\partial}{\partial x}\ln X_q. \quad (43)$$

### 4. The generalized magnetization and susceptibility of Landau Diamagnetism

Materials are generally classified by the sign and the magnitude of their susceptibilities. The majority of the materials (solids) are diamagnetic for which susceptibility $\chi$ is negative. The generalized magnetization $M_q$ in terms of the free energy is given by Eq.(16). On the other hand, when the relation between $F_q$ and $Z_q$ is taken into account the generalized magnetization follows:

$$M_q = n \frac{\hbar e}{mc} Z_q^{1-q} \frac{\partial}{\partial x} \ln Z_q \qquad (44)$$

since

$$\frac{\partial x}{\partial H} = \beta \frac{e\hbar}{mc} \qquad (45)$$

and

$$\frac{\partial F_q}{\partial x} = -\frac{m}{\beta} Z^{1-q} \frac{\partial}{\partial x} \ln Z_q. \qquad (46)$$

Substituting the derivative of $Z_q$ with respect to x which is given by Eq.(43) into Eq.(46) one obtains for the generalized magnetization

$$M_q = n \frac{\hbar e}{mc} Z^{1-q} \left[ \frac{1}{x} - \left(\frac{3-q}{2}\right) \frac{\sum_{n=0}^{\infty}\left(n+\frac{1}{2}\right)\left[1-(1-q)x\left(n+\frac{1}{2}\right)\right]^{\frac{1}{1-q}-\frac{1}{2}}}{X_q} \right] \qquad (47)$$

which is is the generalized exact magnetization expression for the diamagnet, since

$$\frac{\partial}{\partial x} \ln Z_q = -\frac{(3-q)}{2} \sum_{n=0}^{\infty}\left(n+\frac{1}{2}\right)\left[1-(1-q)x\left(n+\frac{1}{2}\right)\right]^{\frac{1}{1-q}-\frac{1}{2}} \bigg/ X_q. \qquad (48)$$

In the high temperature limit, it can be shown that Eq.(41) and Eq.(42) could may take approximated values, i.e.

$$X_q \cong \frac{Y_q}{1-Y_q^2}; \qquad (49)$$

and

$$\frac{\partial}{\partial x}\ln X_q \cong \left(\frac{1+Y_q}{1-Y_q^2}\right)Q_q \tag{50}$$

where

$$Y_q = \left[1-(1-q)\frac{x}{2}\right]^{\frac{1}{1-q}+\frac{1}{2}}; \quad Y_q^2 = [1-(1-q)x]^{\frac{1}{1-q}+\frac{1}{2}}, \quad Q_q = \frac{\partial}{\partial x}\ln Y_q. \tag{51}$$

Thus, the approximated value of magnetization which given by Eq.(47) is obtained in terms of $Y_q$

$$M_q = \frac{n\hbar e}{mc}\left[\frac{1}{x}+\left(\frac{1+Y_q^2}{1-Y_q^2}\right)Q_q\right]Z^{1-q} \tag{52}$$

$$= \frac{n\hbar e}{mc}\left[\frac{1}{x}-\left(\frac{1}{1-q}+\frac{1}{2}\right)\left(\frac{1-q}{2}\right)\frac{1}{\left[1-(1-q)\frac{x}{2}\right]}\frac{1+Y_q^2}{1-Y_q^2}\right]Z^{1-q} \tag{53}$$

since

$$Q_q = -\left(\frac{1}{1-q}+\frac{1}{2}\right)\left(\frac{1-q}{2}\right)\frac{1}{\left[1-(1-q)\frac{x}{2}\right]}. \tag{54}$$

The magnetization is found to be a standard one when $q \to 1$ i.e

$$M_1 = \frac{n\hbar e}{mc}\left[\frac{1}{x}-\frac{1}{2}\coth\frac{x}{2}\right] \tag{55}$$

since

$$Q_q \mid_{q=1} = -\frac{1}{2}; \quad Y_q \mid_{q=1} = \exp(-\frac{x}{2}); \quad Z^{q-1}\mid_{q\to 1}=1 \tag{56}$$

$$\coth\frac{x}{2} = \frac{e^{\frac{x}{2}}+e^{-\frac{x}{2}}}{e^{\frac{x}{2}}-e^{-\frac{x}{2}}}. \tag{57}$$

We keep only the lowest-order contribution of x for $M_1$ in Eq.(55) then the usual expression is obtained

$$M_1 = \frac{n\hbar e}{12mc} x. \tag{58}$$

Now let us consider the diamagnetic susceptibility:

In accordance with Eq.(17) the magnetic susceptibility per electron is given by

$$\chi_q = -\frac{1}{n}\beta \frac{\hbar e}{mc}\frac{\partial M_q}{\partial x} \tag{59}$$

since $\frac{\partial x}{\partial H} = \beta \frac{e\hbar}{mc}$ in accordance with Eq.(45). Substitution Eq.(44) for $M_q$ into Eq.(59) leads to

$$\chi_q = -\beta\left(\frac{\hbar e}{mc}\right)^2 \frac{\partial}{\partial x}\left\{\left[\frac{1}{x}+\frac{\partial}{\partial x}X_q\right]Z^{1-q}\right\} \tag{60}$$

where Eq.(43) is taken into account. Rearranging, then it becomes

$$\chi_q = -\beta\left(\frac{\hbar e}{mc}\right)^2\left[-(2-q)\frac{1}{x^2}+\frac{\partial^2}{\partial x^2}\ln X_q - 2(1-q)\frac{1}{x}\frac{\partial}{\partial x}\ln X_q - (1-q)\left(\frac{\partial}{\partial x}\ln X_q\right)^2\right]Z^{1-q}$$

$$\tag{61}$$

Under the approximation, which is made for diamagnetization, and along the line that $M_q$ is derived, susceptibility is obtained in terms of $Y_q$

$$\chi_q = -\beta\left(\frac{\hbar e}{mc}\right)^2\left[(q-2)\frac{1}{x^2}+\frac{4Y_q^2}{1-Y_q^2}Q_q^2+\left(\frac{1+Y_q^2}{1-Y_q^2}\right)\frac{\partial^2}{\partial x^2}\ln Y_q +\right.$$

$$\left. 2(q-1)\frac{1}{x}\left(\frac{1+Y_q^2}{1-Y_q^2}\right)Q_q + (q-1)\left(\frac{1+Y_q^2}{1-Y_q^2}\right)Q_q^2\right]Z^{1-q}. \tag{62}$$

It is easy to verify that in the q→1 limit, Eq.(62) leads to the standard diamagnetic susceptibility.

$$\chi_1 = -\beta \left(\frac{\hbar e}{mc}\right)^2 \left[-\frac{1}{x^2} + \frac{Y_1^2}{\left(1-Y_1^2\right)^2}\right] \tag{63}$$

or

$$\chi_1 = -\beta \left(\frac{\hbar e}{mc}\right)^2 \left[-\frac{1}{x^2} + \frac{1}{4}\operatorname{csc} h^2 \frac{x}{2}\right]. \tag{64}$$

since

$$Y_1^2 = e^{-x}; \qquad \operatorname{csc} h \frac{x}{2} = \frac{2}{e^{\frac{x}{2}} - e^{-\frac{x}{2}}} \quad . \tag{65}$$

As in the magnetisation one may keep the lowest order contribution in x for susceptibility then Eq.(58) becomes the well-known standard expression [28]:

$$\chi_1 = -\frac{\beta}{3}\left(\frac{e\hbar}{2mc}\right)^2 . \tag{66}$$

**Conclusions**

As is stressed in the introduction BG statistical mechanics is based on the extensive (additive) entropy that optimisation of entropy under the constraints yields an exponential dependence on energy. We study Landau Diamagnetism within the context of generalised TT in which long-range interactions, long range memory multifractal space-time and so forth, i.e. nonextensivity is taken into considerations. In TT formalism entropic index q play a major role and it is an indicator of nonextensivity.

In an earlier paper [21] Landau diamagnetism was considered in the limit where factorisation approximation [15] is valid. Here, we employed the Hillhorst integral transformation in order to obtain the generalised partition functions $Z_q$ exactly, then magnetisation free energy $F_q$, generalised magnetization, $M_q$ and finally generalized susceptibility $\chi_q$ are calculated, high temperature limits of those quantities are also discussed.

In the q→1 limit, standard usual expressions for $Z_1, F_1, M_1, \chi_1$ are also indicated.

According to our justification this paper is on the line to obtain the exact solutions for the nonextensive thermostatistical systems, one of them is Landau diamagnetism.

**Acknowledgement**

FB and DD would like to thank Ege University Research Fund for their partial support under the Project Numbers 98 FEN 25.

**References:**

[1] C.Tsallis, J.Stav. Phys.52,(1988),479; A regularly updated Bibliograpy on the subject is accessible at http://tsallis.cat.cbpf.br/biblio.htm

[2] E.M.F.Curado, C.Tsallis, J.Phys.A 24, (1991), L69; Corrigendo 24 (1991) 3187; 25 (1992) 2019

[3] A.R.Palastino, A.Plastino, Phys.Lett. A A4, (1990),384

[4] P.A.Alemany, D.H.Zanette, Phys.Rev.E49, (1994) R956; C.Tsallis, S.V.F.Levy, A.M.C.Souza, R.Magnard, Phys.Rev.Lett., 75, (1995) 3589; D.H.Zanette, P.A.Alemany, Phys Rev.Lett., 75, (1995) 366; M.O.Careres, C.E.Budde, Phys. Rev.Lett 77, (1996) 2589; D.H.Zanette, P.A. Alemany, Phys.Rev.Lett 77, (1996) 2590

[5] B.M.Boghosian; Phys.Rev.E53 (1996) 4754.

[6] G.Kaniadakis; A.Lavagno, P.Quarati, Phys.Lett. B319 (1996) 308.

[7] A.Lavagno, G.Kaniadakis, M.R.Monterio, P.Quarati, CTsallis, Astrophys Lett. and Comm. 35 (1998) 449.

[8] C.Tsallis, F.C. Sa Barreto, E.D.Loh, Phys.Rev.E.52 (1995) 447.

[9] A.R.Palastino, A.Plastino, H Vcetih, Phys.Rev.Lett A 207 (1998) 42; U.Tırnaklı, F.Büyükkılıç, D.Demirhan, Physica A 240 (1997) 657; Q.A.Wang, A.L.Mehaute,Phys.Rev.Lett A 237 (1997) 28; A.B.Pinheiro, I Raditi, Phys.Rev.Lett A 242 (1998) 296.; Q.A.Wang, A.Le.Mehaute, Phys.Rev.Lett A 242 (1998) 301; E.K.Lenzi, R.S.Mendes, Phys.Rev.Lett A 250, (1998) 270.

[10] D.F.Torres, H.Vucetich, A.Plastino, Phys.Rev.Lett. 79, (1997) 1588.

[11] A.K.Rajogopal, Phys.Rev.Lett. 76, (1996) 3496.

[12] I.KOponen, Phys.Rev.E 55, (1997) 7759.

[13] M.L.Lyra, C.Tsallis, Phys.Rev.Lett 80, (1998) 53.

[14] Constantino Tsallis, Renio S.Mendes, A.R.Plastino, Physica A 261 (1998) 534;B.H.Lavendo, J.Dunning-Davies,Found. Phys.Lett. 3, (1996) 435; Nature 368 (1994) 284.

Bitt.Lavenda, J.Dunniıng-Davies, M Compiani, Nuovo Cimento B 110, (1995) 433;B.H.Landau, Statistical Physics: A Probabilistic Approach, (Wiley-Intercience, New York, 1991);B.H.Landau, Thermodynamics of Extremes, (Albian Chichester, 1995).

[15] F.Büyükkılıç, D.Demirhan, Phys.Lett A 181 (1993) 24;F.Büyükkılıç, D.Demirhan, A. Güleç, Phys.Lett A 197 (1995) 209.

[16] E.K.Lenzi, L.C.Malacarne, R.S. Mendes, Phys.Rev.Lett. 80 (1998) 218.

[17] A.Plastino, C.Tsallis, J.Phys A 26 ( 1993) L893.

[18] L.R.Evangelista, L.C.Malacarne, R.S.Mendes Physica A 253 (1998) 507.

[19] E.K.Lenzi, L.C. Malacarne, R.S.Mendes, Path Integral and Bloch equation in nonextensive Tsallis Statistics, preprint (1998);R.A.Treumann, Generalized-Lorenziann path integrals Phys.Rev E 57 (1998) 5150

[20] A.K.Rajogopal, R.S.Mendes, E.K.Lenzi, Phys.Rev.Lett. 80 (1998) 3907

[21] S.F.Özeren, U.Tırnaklı, F.Büyükkılıç, and D.Demirhan Euro.Phys.J. B2, (1998) 101 and A. Feldman ,A.H. Kahn ,Phys.Rev.B1,(1970) 4584.

[22] C.Tsallis in " New Trends in Magnetism ,Magnetic Materials and Their Appllications " ed. J.L Moran -Lopez and J.M.Sanchez (Plenum Press, New ork,1994) 451.

[23] D.Prato ,Phys.Lett.A 203 (1994)165

[24] J.P.Keener,Principles of Applied Mathematics (Addison-Wesley,New York,1995) p.261.

[25] F.Büyükkilic,D.Demirhan ,Z.Phys.B99(1995)137.

[26] W.Greiner, L Neise, H Stöcker, Thermodynamics and Statistical Mechanics, (Springer Verlag, 1995) p. 367.

[27] I.S. GrandShteyn, I.M.Ryzhik, Table of Integral Series and Products (Academic Press, New York, 1980 ) p.935.

[28] K.Huang, Statistical Mechanics (John Wiley,New York ,1987) p.259.